\pdfoutput=1

\documentclass[11pt]{article}

\usepackage[final]{acl}

\usepackage{times}
\usepackage{latexsym}

\usepackage[T1]{fontenc}

\usepackage[utf8]{inputenc}

\usepackage{microtype}

\usepackage{inconsolata}

\usepackage{graphicx}

\usepackage{multirow}
\usepackage{multicol}
\usepackage{longtable}
\usepackage{graphicx}
\usepackage{array}
\usepackage{colortbl}
\usepackage{amsmath}
\usepackage{booktabs}
\usepackage{arydshln}
\usepackage{amssymb}
\usepackage{subfigure}

\usepackage{xcolor}
\usepackage{listings}

\definecolor{codegreen}{rgb}{0,0.6,0}
\definecolor{codegray}{rgb}{0.5,0.5,0.5}
\definecolor{codepurple}{rgb}{0.58,0,0.82}
\definecolor{backcolour}{rgb}{0.95,0.95,0.92}

\lstdefinestyle{mystyle}{
    backgroundcolor=\color{backcolour},   
    commentstyle=\color{codegreen},
    keywordstyle=\color{magenta},
    numberstyle=\tiny\color{codegray},
    stringstyle=\color{codepurple},
    basicstyle=\ttfamily\footnotesize,
    breakatwhitespace=false,         
    breaklines=true,                 
    captionpos=b,                    
    keepspaces=true,                 
    numbers=left,                    
    numbersep=5pt,                  
    showspaces=false,                
    showstringspaces=false,
    showtabs=false,                  
    tabsize=2
}
\lstdefinelanguage{MyLang}{
    keywords={Input, Output},
    sensitive=false,
    comment=[l]{//},
    morecomment=[s]{/*}{*/},
    morestring=[b]"
}
\title{Evaluating LLMs in Database Scenarios: A Lifecycle Benchmark \\for Assessing Their Potential in Core Database Tasks}

\author{
    Shunfan Zheng\textsuperscript{\rm 1}, Dongsheng Shi\textsuperscript{\rm 1}, Yue Li\textsuperscript{\rm 1}, Xin Yi\textsuperscript{\rm 1}, \textbf{Linlin Wang$^{1}$\thanks{Corresponding Author.}}, \textbf{Gerard de Melo}\textsuperscript{\rm 2}\\ 
    $^{1}$East China Normal University\\  
    $^{2}$Hasso Plattner Institute/University of Potsdam\\  
    \texttt{\{sfzheng, dongsheng, yue\_li, xinyi\}@stu.ecnu.edu.cn,} \\ \texttt{llwang@cs.ecnu.edu.cn, gdm@demelo.org} \\ 
}  
\begin{document}
\maketitle

\begin{abstract}
Large Language Models (LLMs) are transforming database interaction paradigms, evolving from simple query translators to autonomous database administrators (DBAs). However, current evaluation benchmarks remain disproportionately fixated on Text-to-SQL tasks, neglecting the holistic \textbf{Database Lifecycle}-from initial schema design to post-deployment maintenance. This narrow focus fails to capture the diverse capabilities required for real-world database management. To bridge this gap, we introduce \textbf{DBLifeBench}, the first benchmark to evaluate LLMs across five critical lifecycle phases: Design, Implementation, Operation, Debugging, and Maintenance.
Furthermore, addressing the cognitive mismatch between ambiguous natural language and complex SQL logic, we propose \textbf{Progressive-Text2SQL}, a novel task utilizing structured reasoning graphs to mimic human iterative problem-solving. Our extensive evaluation reveals a critical insight: while general-purpose models demonstrate balanced performance, specialized Text-to-SQL models suffer from ``catastrophic forgetting'' in non-coding phases like design and maintenance. DBLifeBench serves as a foundational step toward evaluating and building true full-stack database intelligence.
\end{abstract}

\section{Introduction}
Large Language Models (LLMs) have demonstrated remarkable capabilities in processing structured data, positioning them as potential successors to traditional database interfaces \cite{ruan2023tptu,kong2023tptuv2boostingtaskplanning}. 
Beyond simple query generation, LLMs have increasingly exhibited the potential to serve as autonomous agents capable of planning and executing complex tasks \cite{shi2026surgent, chen2026reinforcement, zhang2026tools}, including managing the entire database pipeline \cite{zhang2026surveyevaluatingqualitytrustworthiness,llm_db_manage,llm_db_2}.
However, as LLMs are increasingly deployed in complex, high-stakes environments, the need for a rigorous, multifaceted evaluation framework has become paramount. High-quality benchmarks serve not only as evaluation tools but as a strategic ``moat'' that drives the evolution of model capabilities, providing the reference baseline essential for building robust database products \cite{schmidt2025sqlstorm}.

\begin{figure}[t]
    \centering
    \includegraphics[width=1\linewidth]{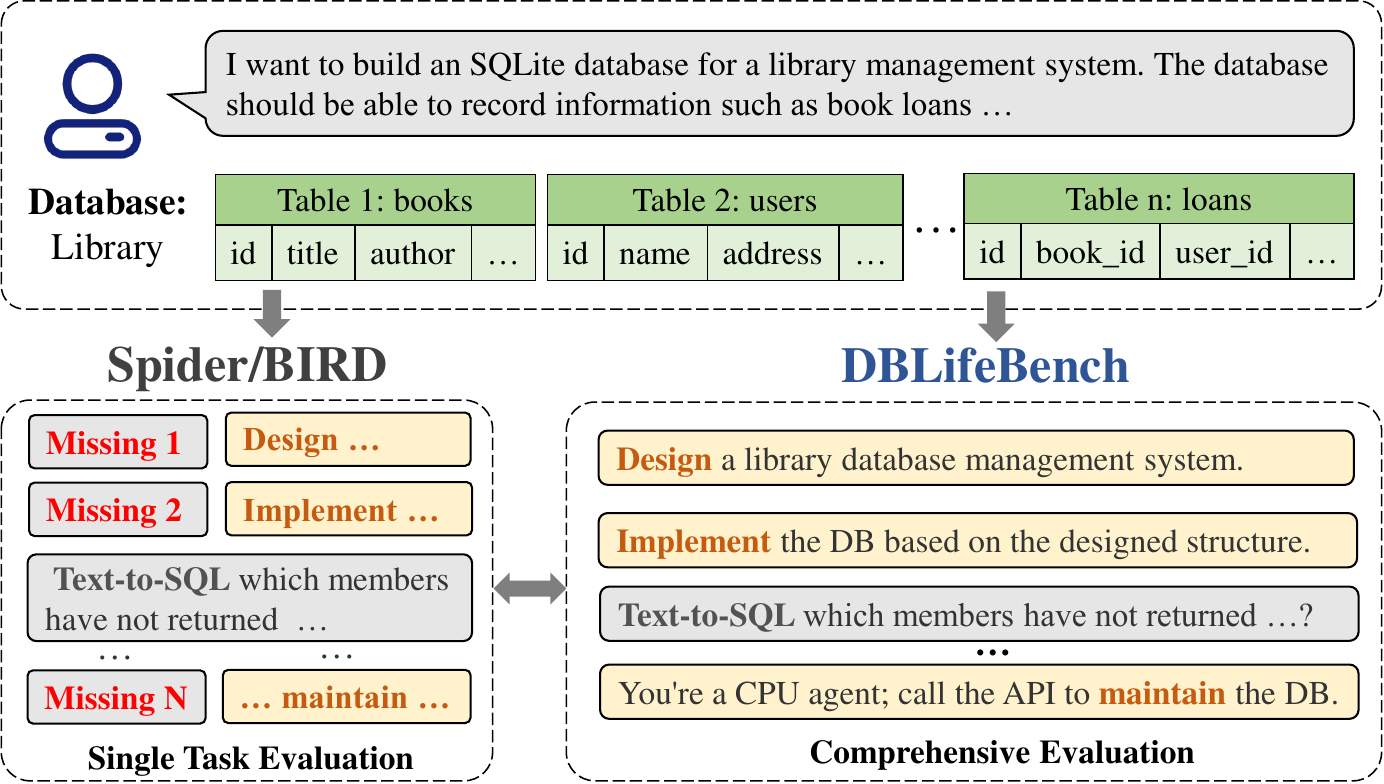}
    \caption{Comparison between traditional Text-to-SQL-centric benchmarks and our holistic DBLifeBench. While existing benchmarks focus solely on query translation, DBLifeBench covers the full spectrum of a DBA's workflow.}
    \label{fig:intro}
\end{figure}

Unfortunately, current benchmarks fail to fulfill this role. Benchmarks such as Spider \cite{spider} and BIRD \cite{BIRD} predominantly focus on the Text-to-SQL task \cite{text2sql_survey,SQLizer,xu2017sqlnet,zelle1996learning}. This narrow focus renders them insufficient as reference baselines for enhancing real-world database product capabilities. While valuable, relying solely on them creates a ``tunnel vision'' effect. Consider a real-world scenario: a database administrator (DBA) for a library system must first \textit{design} a normalized schema to avoid redundancy, \textit{implement} it with correct constraints, and continuously \textit{maintain} it by analyzing error logs. A model that excels at generating SQL SELECT statements (Text-to-SQL) but fails to design a viable schema or debug a deadlock is insufficient for real-world deployment. Optimizing models purely for existing benchmarks yields diminishing returns for actual product utility, as metrics like Execution Accuracy (EX) fail to assess structural and operational reasoning capabilities.

To address these limitations, we introduce \textbf{DBLifeBench}, a comprehensive benchmark designed to evaluate the ``Full-stack Database Intelligence'' of LLMs. DBLifeBench extends beyond query translation to cover five key phases of the database lifecycle: design, implementation, operation, debugging, and maintenance. Each phase features tailored tasks and metrics, ensuring a multidimensional assessment of whether a model can truly function as a database expert.

\begin{figure*}[t]
    \centering
    \includegraphics[width=\linewidth]{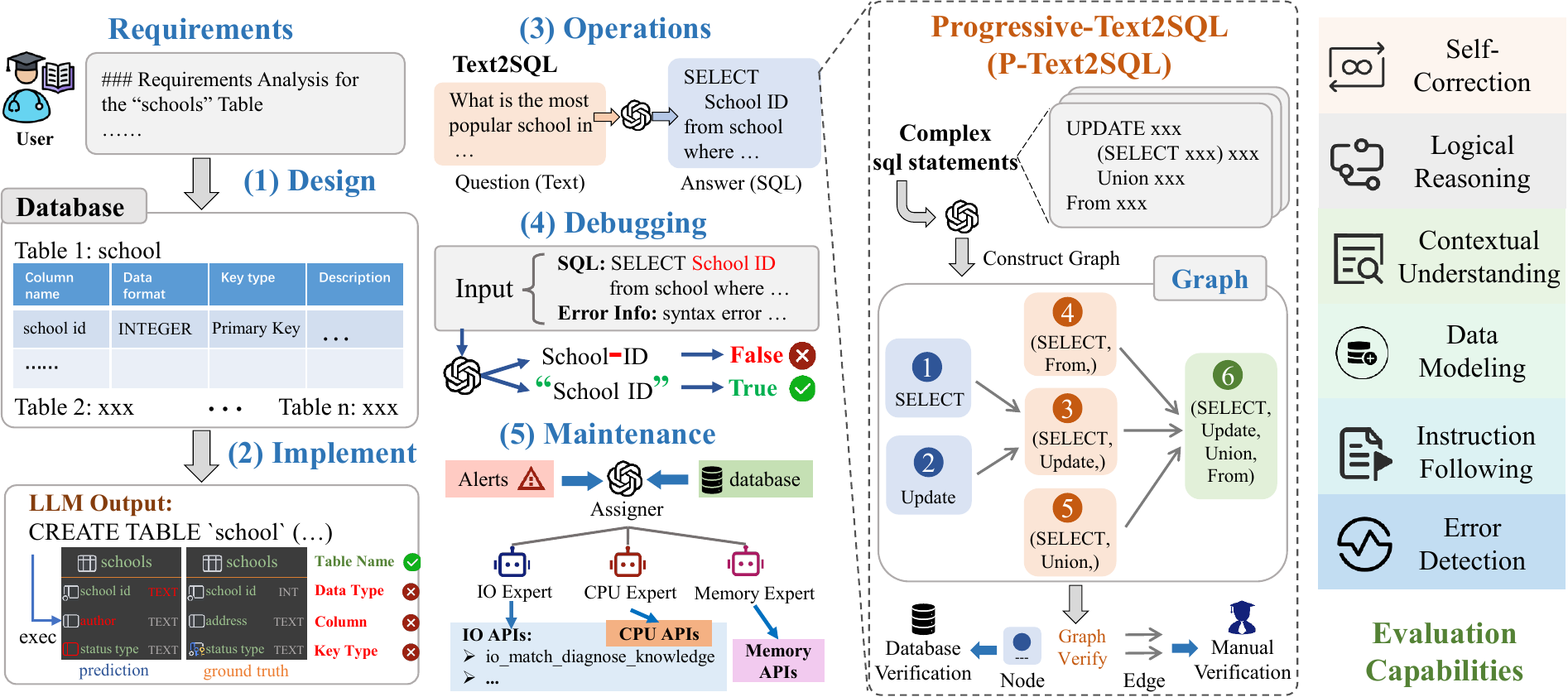}
    \caption{An Overview of DBLifeBench. The left part illustrates the lifecycle-based evaluation, while the right part details the data construction process for the Progressive-Text2SQL task.}
    \label{fig:framework}
    \vspace{-0.5em}
\end{figure*}

Moreover, within the Operation phase, we identify a fundamental cognitive gap in traditional Text-to-SQL: complex SQL logic is often too intricate to be mapped directly from a single natural language sentence. To mitigate this, we propose \textbf{Progressive-Text2SQL}. This task utilizes a Dynamic Reasoning Graph to decompose complex queries into intermediate logical steps, effectively simulating the iterative reasoning process of human experts. This structure not only improves model performance but also aligns evaluations with genuine cognitive workflows.

In summary, our contributions are the follows:
\begin{itemize}
    \item We propose DBLifeBench, the first benchmark to evaluate LLMs across five critical phases of the database lifecycle, moving beyond the single-task paradigm. 
    \item Specifically, we introduce Progressive-Text2SQL, a task that leverages reasoning graphs to bridge the gap between natural language ambiguity and SQL complexity, offering a more robust evaluation of reasoning capabilities.
    \item Our comprehensive analysis reveals the phenomenon of the ``curse of specialization'', whereby models fine-tuned specifically for SQL generation exhibit significant performance degradation in broader database management tasks.
\end{itemize}

\section{The Proposed Benchmark}
\subsection{The Composition of DBLifeBench}
\label{sec:bench_composition}
To comprehensively assess Database Intelligence, DBLifeBench maps the database lifecycle to distinct cognitive capabilities of LLMs. We devise assessments across five phases:

\paragraph{Design (\textit{P}1: Abstraction \& Modeling)} 
This phase evaluates the model's ability to translate abstract requirements into structured schemas. Given natural language requirements \( Q \), the model must construct a schema \( S = ( \mathcal{E}, \mathcal{D}, \mathcal{K} ) \), comprising entities \( \mathcal{E} \), data types \( \mathcal{D} \), and foreign key dependencies \( \mathcal{K} \). This tests the model's high-level logical modeling capabilities.

\paragraph{Implementation (\textit{P}2: Syntax Precision)} 
Models must translate the conceptual schema \( S \) into executable SQL Data Definition Language (DDL) statements \( \mathcal{Y} \). Unlike semantic generation, this phase demands strict syntactic precision to define tables, columns, and constraints that are valid within a specific database engine (e.g., SQLite).

\paragraph{Operation (\textit{P}3: Logic Translation \& Reasoning)} 
This phase focuses on converting operational needs into Data Manipulation Language (DML). It includes standard \textbf{Text2SQL} and our novel \textbf{Progressive-Text2SQL}. For a query or graph node \( Q_i \), the model generates satisfying SQL \( \mathcal{Y}_i \). This tests the model's ability to map natural language intent to database logic.

\paragraph{Debugging (\textit{P}4: Diagnosis \& Correction)}
Simulating a developer's workflow, the model receives a flawed SQL statement \( \mathcal{Y}_{\text{flawed}} \) and an error message. It must identify the root cause-whether syntactic or semantic and generate a corrected statement \( \mathcal{Y}_{\text{corrected}} \).

\paragraph{Maintenance (\textit{P}5: System Management)}
This phase simulates the role of a Site Reliability Engineer (SRE) or DBA. It consists of: (i) \textit{Triage (Assigner)}: Analyzing error logs to assign specific experts (e.g., IOExpert); (ii) \textit{Resolution (Expert)}: Utilizing simulated API tools (e.g., \texttt{check\_metric}) to diagnose and fix system anomalies. This tests the model's domain knowledge and tool-use capabilities.
Figure \ref{fig:example} provides specific examples.

\subsection{Benchmark Construction}
\subsubsection{Overview}
DBLifeBench integrates diverse data sources, including: \textit{S}\#1, existing open-source datasets; \textit{S}\#2, manually constructed data; \textit{S}\#3, authoritative textbooks, and \textit{S}\#4, automatically generated data. 
The entire corpus includes 13 distinct databases, with an average of 7.8 tables per database, covering diverse domains such as finance, education, and sports. We utilize this corpus to  construct five datasets (shown in Table \ref{tab:data_info}) 
for the five core database scenarios introduced in Section \ref{sec:bench_composition}.  
Specific examples are provided in Appendix \ref{sec:appendix_example}. 
\begin{table}[htbp]
    \centering
    \small
    \begin{tabular}{llrc}
    \toprule
    Phase & Dataset & \multicolumn{1}{r}{\# Examples} & Sources \\
    \midrule
    \textit{P}1    & \multirow{2}[0]{*}{Requirements} & \multirow{2}[0]{*}{265} & \multirow{2}[0]{*}{\textit{S}\#1; \textit{S}\#2} \\
    \textit{P}2    &       &       &  \\
    \hdashline
    \multirow{2}[0]{*}{\textit{P}3} & Text2SQL & 1,534 & \textit{S}\#1; \textit{S}\#3 \\
          & P-Text2SQL & 1,149 & \textit{S}\#1; \textit{S}\#3; \textit{S}\#4 \\
    \hdashline
    \textit{P}4    & SQL-Debugging & 523   & \textit{S}\#1; \textit{S}\#2; \textit{S}\#4 \\
    \textit{P}5    & Maintenance & 518   & \textit{S}\#1 \\
    \bottomrule
    \end{tabular}%
    \caption{Dataset Statistics}
    \label{tab:data_info}%
    \vspace{-0.5em}
\end{table}%

\paragraph{Requirements} To ensure high-quality design prompts, we employed a human-AI collaborative loop. The process starts with AI reviewing the schema and generating a description. Three students then draft the initial requirements analysis based on the schema and AI description. AI analyzes this version, suggests improvements, and the students revise it. This iterative process is repeated for three rounds.
\paragraph{Text-to-SQL \& Progressive-Text2SQL} We incorporate BIRD \cite{BIRD} for standard evaluation and construct a new Progressive-Text2SQL dataset to address the ``reasoning hop'' challenges found in complex queries. Using the Dynamic Progressive Reasoning Graph method, we dynamically construct progressive reasoning graphs, detailed in Section \ref{sec:P-Text2SQL}. To ensure data diversity, we also manually include other statements like UPDATE and DELETE, while prior work focuses on SELECT \cite{yu2019cosql,spider2}. 

\paragraph{SQL-Debugging} Consulting BigTable \cite{zhang2024benchmarking}, we collect erroneous SQL statements and their corresponding error messages during the execution of Text-to-SQL tasks. Based on this, we construct the SQL Debugging dataset. These erroneous SQL statements cannot be executed properly in SQLite and require correction.

\paragraph{Maintenance} Based on the DB-GPT framework \cite{db_gpt}, we divide tasks into assignment and expert execution. Detailed examples are provided in Figure \ref{fig:example}. Automated techniques like regular expression matching and database entity recognition were used for filtering, followed by execution and validation in a database environment to ensure effectiveness.



\subsubsection{Progressive-Text2SQL}
\label{sec:P-Text2SQL}
Standard Text-to-SQL tasks often suffer from a misalignment between the brevity of natural language and the complexity of the target SQL. To address this, Progressive-Text2SQL transforms the task into a structured reasoning process (Figure \ref{fig:framework}, right).

\paragraph{SQL Reasoning Graph}
We define a reasoning graph \(G = (V, E)\) to serve as a cognitive scaffold for the model. Nodes \(V = \{v_1, \dots, v_n\}\) represent sub-tasks (e.g., ``filter by date'' or ``join tables''), where each node contains a localized natural language description and its corresponding SQL fragment. Edges \(E\) represent logical dependencies, guiding the model to build the final query \(v_n\) incrementally. This structure mimics the ``Chain-of-Thought (CoT)'' process of human experts.


\paragraph{Graph Construction}
The data construction process is divided into three parts: 
\textit{(i)} We collect examples from Text2SQL that are more difficult and where the misalignment between the natural language description and SQL has been manually identified.
\textit{(ii)} To mitigate potential bias from relying on a single model, we construct the reasoning graph using a diverse set of large language models, including GPT-4o, Claude, Gemini, and DeepSeek-R1. Each model independently generates candidate graph structures based on complex SQL queries. We also provide a few manually written examples in the prompt to assist the LLM in understanding and generating the graph. The specific process is detailed in Appendix \ref{apd:p-text2sql}.
\textit{(iii)} Based on the constructed SQL reasoning graph, we then use the LLM to generate the corresponding natural language description for each SQL node in the graph. The natural language description and SQL from the original example are provided in the prompt. 


\paragraph{Data Validation}
We employ a two-stage validation strategy. First, all SQL nodes are executed in an SQLite environment, and those failing to execute are discarded. Following this, three graduate students with backgrounds in database systems independently reconstruct graph edges using only the SQL and natural language descriptions for each node, without seeing the original model-generated edges. We then compare their annotations with the model-generated structure using Jaccard similarity. Only examples achieving more than 80\% consensus are retained. To quantify inter-rater reliability \cite{inter_agreement}, we calculate Fleiss' kappa, obtaining a score of 0.8173, which indicates strong agreement. Annotators are trained with representative edge construction examples and participate in weekly calibration discussions. Additionally, all annotation disagreements are logged and analyzed to identify common error patterns such as alias confusion and aggregation mismatches, which in turn inform future refinements to the data construction pipeline.

\section{Evaluation Metrics}

As shown in Section \ref{sec:bench_composition}, the evaluations in DBLifeBench vary across different phases, necessitating a tailored set of assessment metrics based on the unique characteristics of each task.

(1) For the design phase: we assess whether the model's designed database structure is reasonable in this phase. We establish three evaluation metrics: entity name accuracy (ACC$_i$-Entity), entity data type accuracy (ACC$_i$-Data), and foreign key dependency accuracy (ACC$_i$-Key). $i$ denotes the number of output tables, defaulting to 2. ACC$_i$-Entity can computed by:
\begin{equation}
\begin{aligned}
\text{ACC$_i$-Entity} &= \frac{\sum_{t=1}^{T} \left| E_t \cap \hat{E}_t \right|}{\sum_{t=1}^{T} \left| E_t \right|}\text{,}\\
\end{aligned}
\end{equation}
where \( T \) is the total number of tables, \( \hat{E_t} \) represents the ground-truth set of entity names for table \( t \), and \( E_t \) represents the set of entity names predicted by the model for table \( t \). Analogously, the same approach is followed for the other metrics.

\begin{table*}[h]
    \centering
    \small
    \resizebox{0.95\textwidth}{!}{
    \begin{tabular}{>{\centering\arraybackslash}m{0.5cm}|lccccccc}
        \toprule
        \multicolumn{2}{c}{\multirow{2}[0]{*}{Model}} & \multicolumn{4}{c}{Design}    & \multicolumn{3}{c}{Implementation} \\
        
        \cmidrule(lr){3-6}\cmidrule(lr){7-9}
        
        \multicolumn{2}{c}{} & \multicolumn{1}{l}{ACC$_2$-Entity} & \multicolumn{1}{c}{ACC$_i$-Data} & \multicolumn{1}{c}{ACC$_i$-Key} & \multicolumn{1}{c}{AVG} & \multicolumn{1}{c}{T-level} & \multicolumn{1}{c}{F-level} & \multicolumn{1}{c}{AVG} \\
        
        \midrule
        \multirow{7}[0]{*}{\rotatebox{90}{General}} 
              & GPT-4o & \textbf{76.94} & \textbf{65.97} & \textbf{80.99} & \textbf{74.63} & \textbf{85.71} & \textbf{77.35} & \textbf{81.53} \\
              & GPT-4o-mini & 69.30 & 55.84 & 79.95 & 68.36 & 79.03 & 71.30 & 75.17 \\
              & Llama3 & 64.90 & 50.20 & 73.14 & 62.74 & 70.41 & 68.59 & 71.00 \\
              & Mistral & 41.99 & 31.72 & 39.16 & 37.62 & 56.08 & 39.23 & 47.66 \\
              & DeepSeek & 33.46 & 25.96 & 37.59 & 32.34 & 42.88 & 32.55 & 37.72 \\
              & Qwen2.5  & 66.31 & 53.80 & 67.71 & 62.61 & 80.39 & 65.81 & 73.10 \\
              & ChatGLM-4 & 67.08 & 52.48 & 72.62 & 64.06 & 78.71 & 69.24 & 73.98 \\
        \midrule
        \multirow{4}[0]{*}{\rotatebox{90}{Specialized}} & DeepSeek-Coder & 66.78 & 56.62 & 74.84 & 66.08 & 78.28 & 64.18 & 71.23 \\
              & SQLCoder & 52.79 & 40.60 & 59.19 & 50.86 & 10.00 & 7.35 & 8.68 \\
              & CodeQwen & 69.24 & 59.45 & 75.33 & 68.00 & 73.86 & 61.83 & 67.85  \\
              & Llama3-sqlcoder & 45.95 & 35.45 & 46.40 & 42.60 & 38.45 & 29.17 & 33.81 \\
        \bottomrule
    \end{tabular}%
    }
    \vspace{0.2em}
    \resizebox{0.95\textwidth}{!}{
    \begin{tabular}{>{\centering\arraybackslash}m{0.5cm}|lccccccc}
        \toprule
    
        \multicolumn{2}{c}{\multirow{2}[0]{*}{Model}} & \multicolumn{2}{c}{Operation} & \multicolumn{2}{c}{Debugging} & \multicolumn{3}{c}{Maintenance} \\
        
        \cmidrule(lr){3-4}\cmidrule(lr){5-6}\cmidrule(lr){7-9}

        \multicolumn{2}{c}{} & \multicolumn{1}{c}{Text2SQL} & \multicolumn{1}{c}{P-Text2SQL} & \multicolumn{1}{c}{Single-round} & \multicolumn{1}{c}{Multi-round} & \multicolumn{1}{c}{Assigner} & \multicolumn{1}{c}{Expert} & \multicolumn{1}{c}{AVG} \\
        \midrule
        \multirow{7}[0]{*}{\rotatebox{90}{General}}
              & GPT-4o & \textbf{56.19} & 53.79 & \textbf{68.39} & \textbf{72.17} & \textbf{79.38} & 33.50 & \textbf{56.44} \\
              & GPT-4o-mini & 47.72 & \textbf{54.31} & 60.44 & 61.83 & 72.88 & \textbf{33.58} & 53.23 \\
              & Llama3 & 30.51 & 45.63 & 38.04 & 43.42 & 59.32 & 32.77 & 61.55 \\
              & Mistral & 19.82 & 28.55 & 23.62 & 36.82 & 46.47 & 3.58 & 25.03 \\
              & DeepSeek & 17.54 & 26.81 & 23.09 & 27.81 & 53.39 & 31.25 & 42.32 \\
              & Qwen2.5  & 38.01 & 39.08 & 35.38 & 42.11 & 67.80 & 32.25 & 50.02 \\
              & ChatGLM-4 & 35.98 & 46.13 & 30.78 & 40.98 & 46.89 & 33.17 & 40.03 \\
        \midrule
        \multirow{4}[0]{*}{\rotatebox{90}{Specialized}} & DeepSeek-Coder & 39.31 & 47.00 & 41.53 & 50.69 & 45.48 & 33.50 & 39.49 \\
              & SQLCoder & 24.64 & 29.85 & 25.74 & 31.89 & 0.00 & 0.00 & 0.00 \\
              & CodeQwen &37.35 & 41.22 & 21.99 & 23.32 & 35.31 & 32.25 & 33.78 \\
              & Llama3-sqlcoder & 33.25 & 39.43 & 48.18 & 62.52 & 46.05 & 32.92 & 39.48 \\
        \bottomrule
    \end{tabular}%
    }
    \caption{Main Results on DBLifeBench}
    \label{tab:main}%
    \vspace{-0.5em}
\end{table*}%

(2) For the implementation phase: in this phase, the model outputs the creation statements for database tables. We execute these statements and evaluate the \(T\) resulting tables using table-level accuracy (T-Level) and field-level accuracy (F-Level):
\begin{equation}
\small
\left\{
\begin{aligned}
\text{T-Level} &= \frac{1}{T}\sum_{t=1}^{T} \mathbb{I}(N_t, \hat{N}_t)\\
\text{F-Level} &= \frac{\sum_{t=1}^{T} \left| F_t \cap \hat{F}_t \right|}{\sum_{t=1}^{T} \left| \hat{F}_t \right|}\\
\end{aligned}
\right.
\end{equation}
Here, \( N_t \) represents the predicted table name for table \( t \) and \( F_t \) represents the set of fields for predicted table \( t \). \(\hat{N}_t\) and \(\hat{F}_t\) represent the corresponding ground truth. Finally, \(\mathbb{I}(\cdot)\) is an indicator function that yields 1 whenever \(Y\) and \(\hat{Y}\) are semantically consistent (not necessarily identical in name), and 0 when \(Y\) and \(\hat{Y}\) are not semantically consistent.

(3) For the operation phase: in the case of the Text2SQL task, we adopt the widely used Execution Accuracy (EX) metric, which calculates accuracy by comparing the execution results \(\hat{V_n}\) of the \(n\)-th predicted SQL statement with the execution results \(V_n\) of the ground truth SQL:
\begin{equation}
\text{EX} = \frac{1}{N}\sum_{n=1}^N \mathbb{I}(V_n, \hat{V}_n)
\end{equation}
For the Progressive-Text2SQL task, we apply EX for each node, and define the Graph Execution Accuracy (G-EX) as the average of the node accuracies, i.e.
\begin{equation}
\text{G-EX} = \frac{1}{G} \sum_{g=1}^{G} \text{EX}_g,
\end{equation}
where \( G \) is the number of reasoning graphs, and \( \text{EX}_g \) represents the Execution Accuracy of the \( g \)-th graph.

(4) For the debugging phase: we execute the corrected SQL generated by the model on the database and evaluate the model's error correction ability based on the execution results.

(5) For the maintenance phase: we evaluate the correctness of expert assignments (by the Assigner) and API calls (by the Expert).

\section{Experiments}
\subsection{Baseline Models}
We evaluate the performance of two classes of advanced baseline models in DBLifeBench. The first class consists of general models, including Open\-AI's GPT series (GPT-4o, GPT-4o-mini), Llama3, Mistral, DeepSeek, Qwen, and ChatGLM-4. The second class comprises specialized models that have been fine-tuned on SQL, code, and related data, including DeepSeek-Coder, SQLCoder, Code\-Qwen, and Llama3-Coder. 

\subsection{Experimental Setup}
We run our extensive experiments using the SQLite database, consistent with previous work such as BIRD \cite{BIRD}. 
During inference, we consistently set the temperature to 0.3 and the top-$p$ to 0.2 to ensure a balance between diversity and coherence in the generated outputs.

\subsection{Main Results}
We conducted a comprehensive evaluation of 11 LLMs using our DBLifeBench, examining their performance across five different phases of the database lifecycle. The experimental results are presented in Table \ref{tab:main}.
GPT-4o and GPT-4o-mini generally outperform other models across almost all aspects, demonstrating their comprehensive database capabilities. However, certain models show varying performance in different phases of evaluation, underscoring the need for a more comprehensive evaluation of different phases. For instance, ChatGLM-4 excels in the Design and Implementation phases, but performs poorly in the Operation and Maintenance phases. Some models, such as SQLCoder, struggle significantly in the Implementation and Maintenance phases, likely due to their training data being heavily focused on Text2SQL tasks, which adversely affects their performance in other areas.

\textbf{Specialized models struggle with ``Full-stack'' tasks.} DeepSeek, after fine-tuning with code data, shows significant improvements across the lifecycle evaluation, while Llama3, fine-tuned on SQL data, experiences a decline in the Design and Implementation phases. Task-specific fine-tuning can sometimes cause models to over-focus on particular patterns, impairing their generalizability to other types of data. In contrast, general-purpose models appear less affected by this overfitting and can better cope with variation. Additionally, the task itself may have scale dependencies: As shown in Section \ref{sec:exp_table_nums}, when the number of input tables is low, the task is simpler, and general models might perform better since specialized models' fine-tuned features may not offer significant advantages.


\textbf{Some models lack the ability to perform multi-round debugging.} Models such as Llama3 and Mistral show significant improvements in multi-round debugging compared to single-round debugging, indicating that they are better at recognizing and correcting their previous mistakes. In contrast, DeepSeek lacks this capability, as its performance in the multi-round setting is consistent with that of single-round debugging, showing no improvement.

\textbf{Model performance is different in the two maintenance tasks.} As Assigners, models like GPT-4o and Qwen are adept at discerning error types from logs, but others are not. This reflects differences in the models' ability to perceive database contexts and the richness of their database knowledge. As Experts, all models exhibit similar performance and obtain low scores because they are not good at using database tools.


\section{Analysis}

\subsection{Impact of Data Model Variations}
To assess the impact of schema variation on LLMs' database performance, we conducted experiments using three schema variants: v1 (lowest normalization), v2, and v3 (highest normalization), all sourced from \cite{furst2024evaluating} and migrated to SQLite. These are evaluated on Text2SQL and P-Text2SQL tasks.

Table \ref{tab:data_model} shows LLM performance across schemas. v2 underperformed due to alias conflicts and redundant UNION operations from bridge tables, which make queries more complex. In contrast, v3 performed better by merging tables, eliminating aliases, and aligning schema semantics with natural language queries. This suggests that over-normalization should be avoided, with a focus on semantic alignment and minimizing multi-table joins. Additionally, as shown in Table \ref{tab:data_model_CV}, the coefficient of variation \cite{CV_reed2002use} is smaller in the P-Text2SQL task, indicating that P-Text2SQL helps the model better adapt to database data model variations.

\begin{table}[t]
  \centering
  \small
    \begin{tabular}{lrr}
    \toprule
    Model & \multicolumn{1}{l}{Text2SQL} & \multicolumn{1}{l}{P-Text2SQL} \\
    \midrule
    GPT-4o & 5.30  & 4.30  \\
    Llama3 & 9.30  & 7.57  \\
    Qwen2.5 & 13.42  & 5.80  \\
    CodeQwen & 13.42  & 5.80  \\
    DeepSeek & 7.91  & 6.83  \\
    DeepSeek-Coder & 11.51 & 5.95 \\
    \bottomrule
    \end{tabular}%
    \caption{Coefficient of Variation}
    \label{tab:data_model_CV}%
    \vspace{-0.5em}
\end{table}%

\subsection{Iteration of Requirement Analysis}
\begin{figure}[t]
    \centering
    \subfigure[Human Assessment]{
        \includegraphics[width=0.46\linewidth]{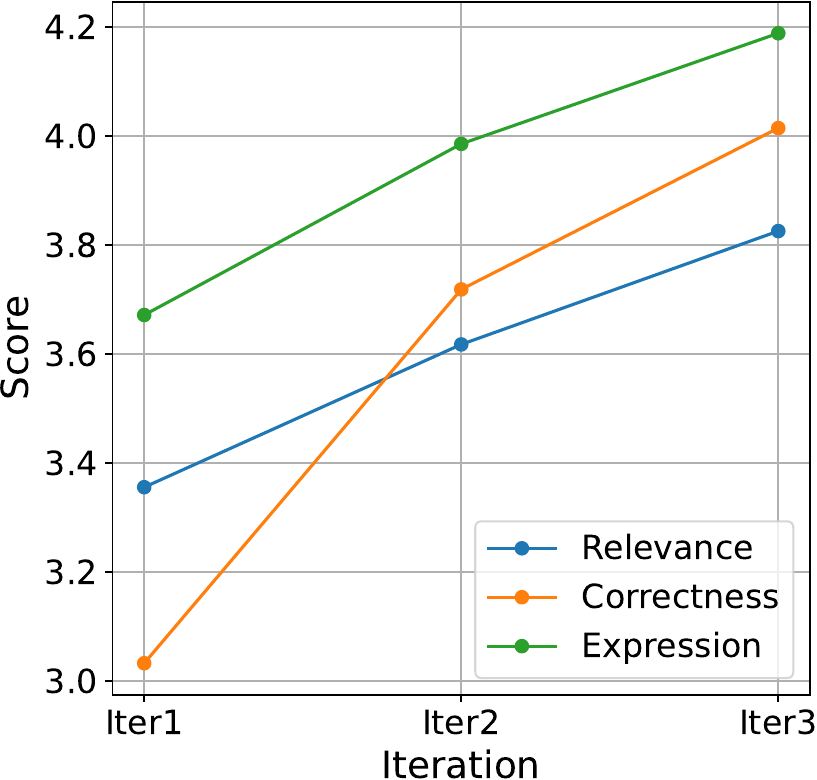}
        \label{fig:req_a}
    }
    \subfigure[Model Performance]{
        \includegraphics[width=0.46\linewidth]{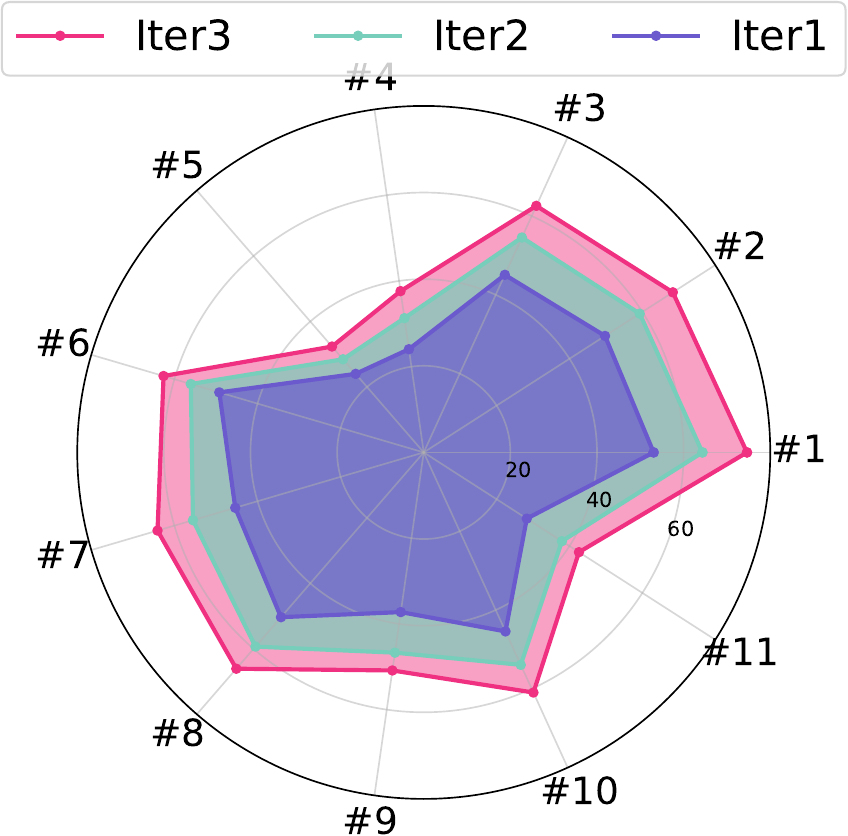}
        \label{fig:req_b}
    }    
    \caption{(a) Human quality assessment of different versions of requirements analysis. (b) Performance of 11 models with different versions of requirements analysis, ordered as in Table \ref{tab:main}.}
    \label{fig:requirement_iter}
    \vspace{-0.5em}
\end{figure}

The requirements analysis dataset underwent three rounds of iterative revisions. To assess the improvements, we analyzed the dataset after each iteration from both human evaluation and model performance perspectives. Figure \ref{fig:req_a} shows that three master's students assessed the dataset on relevance, correctness, and expression. The results indicate significant improvement after each iteration, confirming enhanced clarity and accuracy. Fleiss' kappa coefficient \cite{fleiss1981measurement} yielded a value of 0.8171, indicating strong consistency among annotations. Figure \ref{fig:req_b} shows that 11 baseline models performed better in the database design phase as the quality of the requirements analysis improved, suggesting that higher-quality analysis leads to better model performance and fewer evaluation errors.

\subsection{In-depth Analysis with P-Text2sql}
\begin{figure}[t]
    \centering
    \subfigure[EX Comparison]{
        \includegraphics[width=0.46\linewidth]{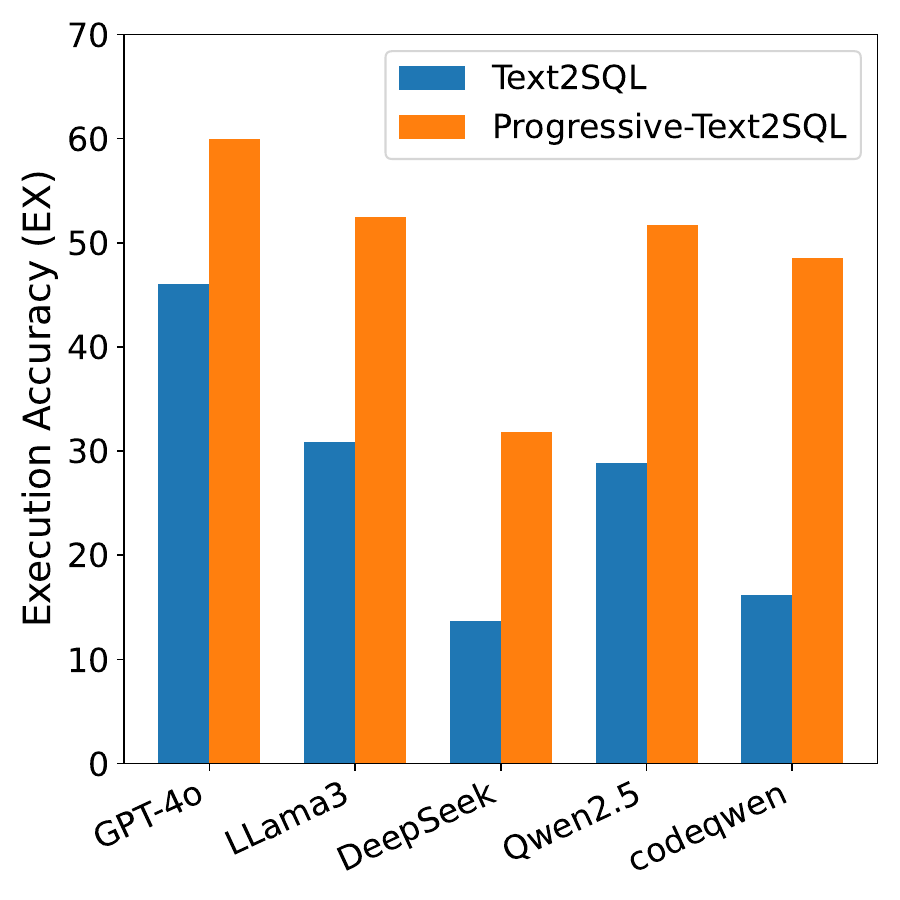}
        \label{fig:text_vs_graph}
    }    
    \subfigure[Growth Rate Comparison]{
        \includegraphics[width=0.46\linewidth]{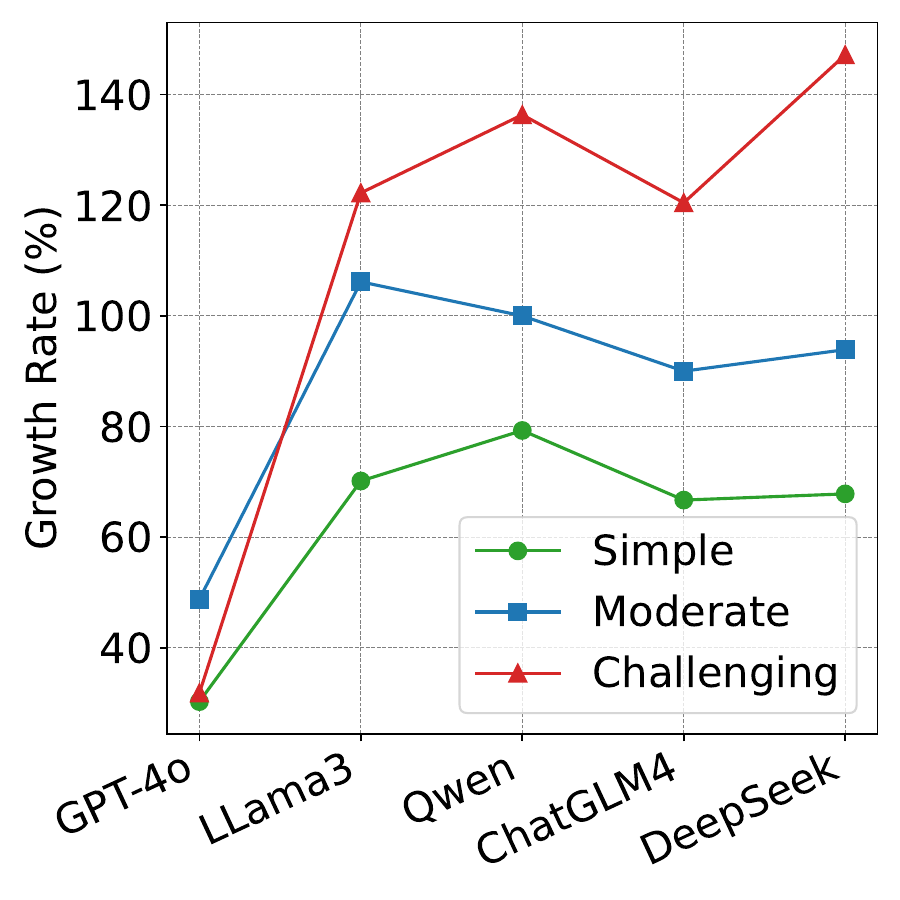}
        \label{fig:text_vs_graph_diff}
    }    
    \caption{Comparison of the model's EX results on Progressive-Text2SQL and Text2SQL tasks. The Progressive-Text2SQL settings slightly differ from the main experiment. We only compute the EX for the final node of each graph to ensure consistency with the Text2SQL data.}
    \vspace{-0.5em}
\end{figure}

To explore whether the dynamic progressive reasoning graph in Progressive-Text2SQL can better assist models in generating complex SQL queries, we compare the performance of models in Text2SQL and Progressive-Text2SQL tasks, as shown in Figure \ref{fig:text_vs_graph}. Graphs significantly improve the Execution Accuracy (EX) of all baseline models, indicating that the graph effectively simulates incremental user input, helping the model better understand and complete SQL generation tasks.

Additionally, we compare the growth rate of EX for SQL generation tasks of varying difficulty when a graph is applied. As shown in Figure \ref{fig:text_vs_graph_diff}, as the task difficulty increases, the improvement in SQL generation due to the use of graphs becomes more pronounced, with the most significant increase observed for tasks at the ``Challenging'' difficulty level. This is because more difficult SQL queries tend to be longer and harder to describe clearly using simple natural language. With graphs, however, the model is guided through a step-by-step reasoning process, helping it generate more accurate SQL queries.

\subsection{Robustness Test of P-Text2SQL}  

In the Progressive-Text2SQL dataset, we constructed high-quality reasoning graphs, but real-world reasoning may contain errors. To better simulate realistic disruptions and evaluate model robustness, we define level-\(k\) perturbations as the targeted selection of nodes within the reasoning graph, where each selected node is subjected to one or more disruptive operations. These operations include syntactic corruption of SQL statements, semantic misalignment with the schema, injection of contradictory logic, or complete removal of the node. This approach introduces realistic flaws that can affect downstream reasoning. As shown in Table \ref{tab:robust_DPRG}, model performance declines as the perturbation level increases, yet in most cases, still outperforms baseline Text2SQL models. This highlights the importance of step-by-step structured reasoning and the critical role of graph integrity in SQL generation tasks.
\begin{figure}[t]
    \centering
    \includegraphics[width=1\linewidth]{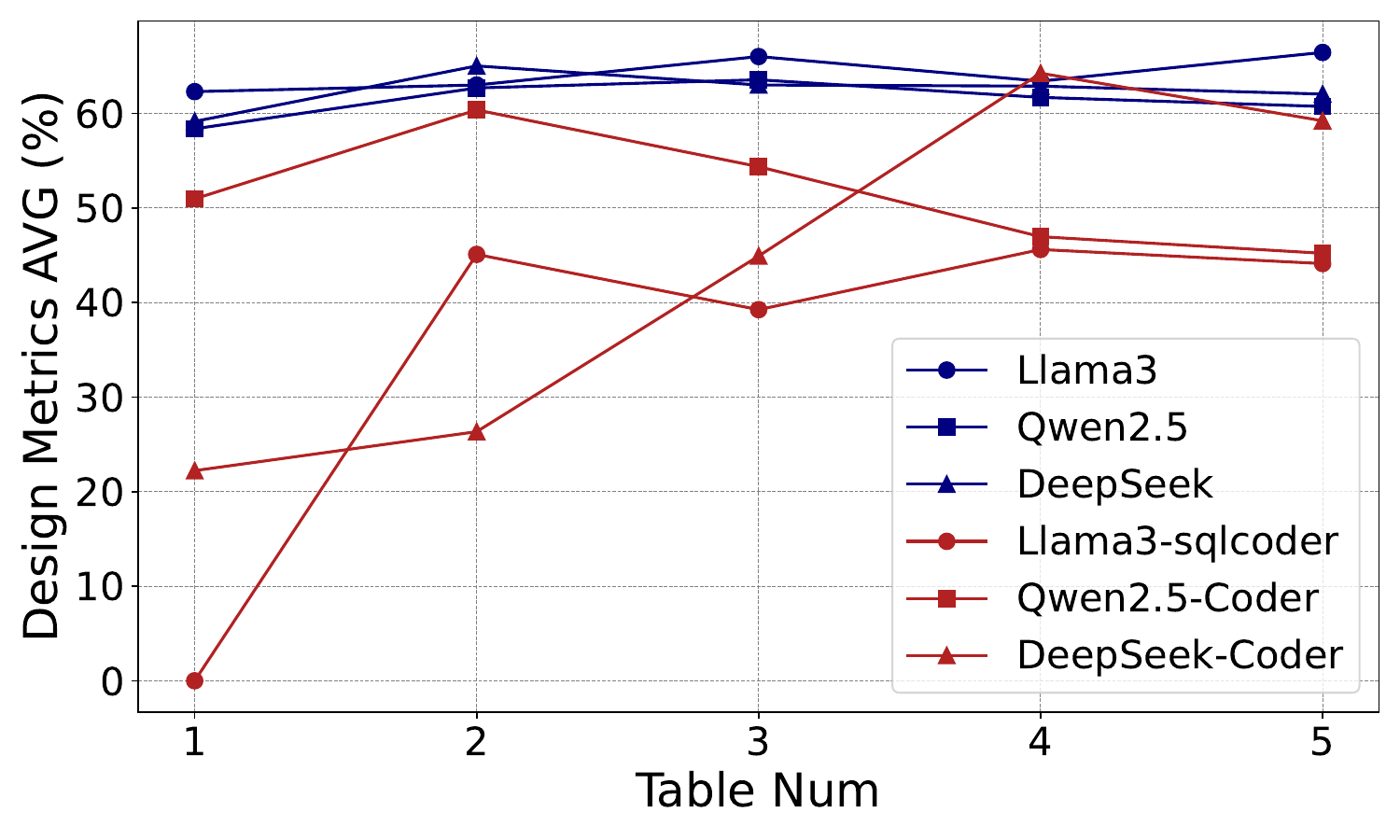}
    \caption{Impact of the number of tables on LLMs in database design tasks.}
    \label{fig:table_num}
    \vspace{-0.5em}
\end{figure}

\begin{table*}[t]
    \centering
    
    \resizebox{1.0\textwidth}{!}{
    \begin{tabular}{lccccccccc}
    \toprule
          & \multicolumn{1}{l}{Llama3} & \multicolumn{1}{l}{Mistral} & \multicolumn{1}{l}{DeepSeek} & \multicolumn{1}{l}{Qwen2.5} & \multicolumn{1}{l}{ChatGLM-4} & \multicolumn{1}{l}{DeepSeek-Coder} & \multicolumn{1}{l}{sqlcoder} & \multicolumn{1}{l}{CodeQwen} & \multicolumn{1}{l}{Llama3-sqlcoder} \\
    \midrule
    P-Text2SQL & \cellcolor[rgb]{ .173,  .345,  .459}46.76 & \cellcolor[rgb]{ .498,  .643,  .69}30.1 & \cellcolor[rgb]{ .537,  .678,  .718}28.11 & \cellcolor[rgb]{ .165,  .337,  .455}47.01 & \cellcolor[rgb]{ .176,  .349,  .463}46.52 & \cellcolor[rgb]{ .18,  .353,  .467}46.27 & \cellcolor[rgb]{ .376,  .529,  .604}36.32 & \cellcolor[rgb]{ .278,  .443,  .537}41.29 & \cellcolor[rgb]{ .224,  .392,  .498}44.03 \\
    Level-1  & \cellcolor[rgb]{ .235,  .4,  .506}43.71 & \cellcolor[rgb]{ .502,  .647,  .694}30.02 & \cellcolor[rgb]{ .576,  .714,  .745}26.24 & \cellcolor[rgb]{ .176,  .349,  .463}46.48 & \cellcolor[rgb]{ .2,  .369,  .482}45.27 & \cellcolor[rgb]{ .231,  .396,  .502}43.81 & \cellcolor[rgb]{ .416,  .565,  .631}34.27 & \cellcolor[rgb]{ .278,  .443,  .537}41.18 & \cellcolor[rgb]{ .259,  .424,  .522}42.39 \\
    Level-2  & \cellcolor[rgb]{ .275,  .435,  .533}41.48 & \cellcolor[rgb]{ .608,  .741,  .769}24.65 & \cellcolor[rgb]{ .639,  .769,  .792}22.89 & \cellcolor[rgb]{ .329,  .486,  .573}38.81 & \cellcolor[rgb]{ .353,  .51,  .588}37.65 & \cellcolor[rgb]{ .302,  .463,  .553}39.98 & \cellcolor[rgb]{ .51,  .651,  .698}29.58 & \cellcolor[rgb]{ .455,  .6,  .659}32.34 & \cellcolor[rgb]{ .369,  .522,  .6}36.90 \\
    Level-3  & \cellcolor[rgb]{ .376,  .529,  .604}36.29 & \cellcolor[rgb]{ .655,  .784,  .804}22.13 & \cellcolor[rgb]{ .769,  .886,  .882}16.36 & \cellcolor[rgb]{ .435,  .584,  .647}33.33 & \cellcolor[rgb]{ .435,  .584,  .647}33.33 & \cellcolor[rgb]{ .38,  .533,  .608}36.07 & \cellcolor[rgb]{ .651,  .78,  .8}22.45 & \cellcolor[rgb]{ .58,  .718,  .749}25.87 & \cellcolor[rgb]{ .494,  .635,  .686}30.35 \\
    Level-4  & \cellcolor[rgb]{ .435,  .584,  .647}33.28 & \cellcolor[rgb]{ .69,  .816,  .827}20.40 & \cellcolor[rgb]{ .765,  .882,  .878}16.67 & \cellcolor[rgb]{ .502,  .647,  .694}29.85 & \cellcolor[rgb]{ .471,  .616,  .671}31.59 & \cellcolor[rgb]{ .431,  .58,  .643}33.68 & \cellcolor[rgb]{ .686,  .812,  .824}20.71 & \cellcolor[rgb]{ .667,  .792,  .808}21.60 & \cellcolor[rgb]{ .553,  .69,  .729}27.36 \\
    Level-5  & \cellcolor[rgb]{ .459,  .608,  .663}32.14 & \cellcolor[rgb]{ .706,  .827,  .835}19.65 & \cellcolor[rgb]{ .773,  .89,  .886}16.23 & \cellcolor[rgb]{ .502,  .647,  .694}29.85 & \cellcolor[rgb]{ .482,  .627,  .682}30.85 & \cellcolor[rgb]{ .451,  .596,  .659}32.68 & \cellcolor[rgb]{ .706,  .827,  .835}19.65 & \cellcolor[rgb]{ .698,  .824,  .831}19.90 & \cellcolor[rgb]{ .569,  .702,  .741}26.59 \\
    Text2SQL & \cellcolor[rgb]{ .522,  .663,  .71}28.85 & \cellcolor[rgb]{ .792,  .91,  .898}15.17 & \cellcolor[rgb]{ .82,  .933,  .918}13.68 & \cellcolor[rgb]{ .522,  .663,  .71}28.86 & \cellcolor[rgb]{ .522,  .663,  .71}28.86 & \cellcolor[rgb]{ .482,  .627,  .682}30.85 & \cellcolor[rgb]{ .765,  .882,  .878}16.67 & \cellcolor[rgb]{ .769,  .886,  .882}16.42 & \cellcolor[rgb]{ .6,  .733,  .765}24.88 \\
    \bottomrule
    \end{tabular}%
    }
  \caption{Robustness Test of Progressive-Text2SQL}
  \label{tab:robust_DPRG}%
  \vspace{-0.5em}
\end{table*}%

\subsection{Influence of Number of Tables for Database Design}
\label{sec:exp_table_nums}
In the main experiment, the design phase evaluation requires the model to design two tables within the database. To further investigate how the number of tables to be designed affects the model, we modify the input to include the description of the entire database, with the model being tasked to output the design for a specified number of tables, ranging from 1 to 5. We then evaluate the design capabilities of models under these different requirements.
As shown in Figure \ref{fig:table_num}, as the number of tables grows, different models exhibit varying performance. Among them, the general-purpose models show relatively stable and consistent performance, with little impact from the number of tables. In contrast, models fine-tuned on code data exhibit more significant fluctuations in performance, which may be due to the impact of code fine-tuning on the general instruction-following capabilities of the model.  Some specialized models, in most cases, perform worse with fewer input tables compared to when they have more tables. This suggests that when the number of input tables is small, specialized models may struggle to fully utilize their fine-tuned features in such simpler tasks.

\subsection{New Perspective of Model Rankings}

\begin{figure}[h]
    \centering
    
    \includegraphics[width=1\linewidth]{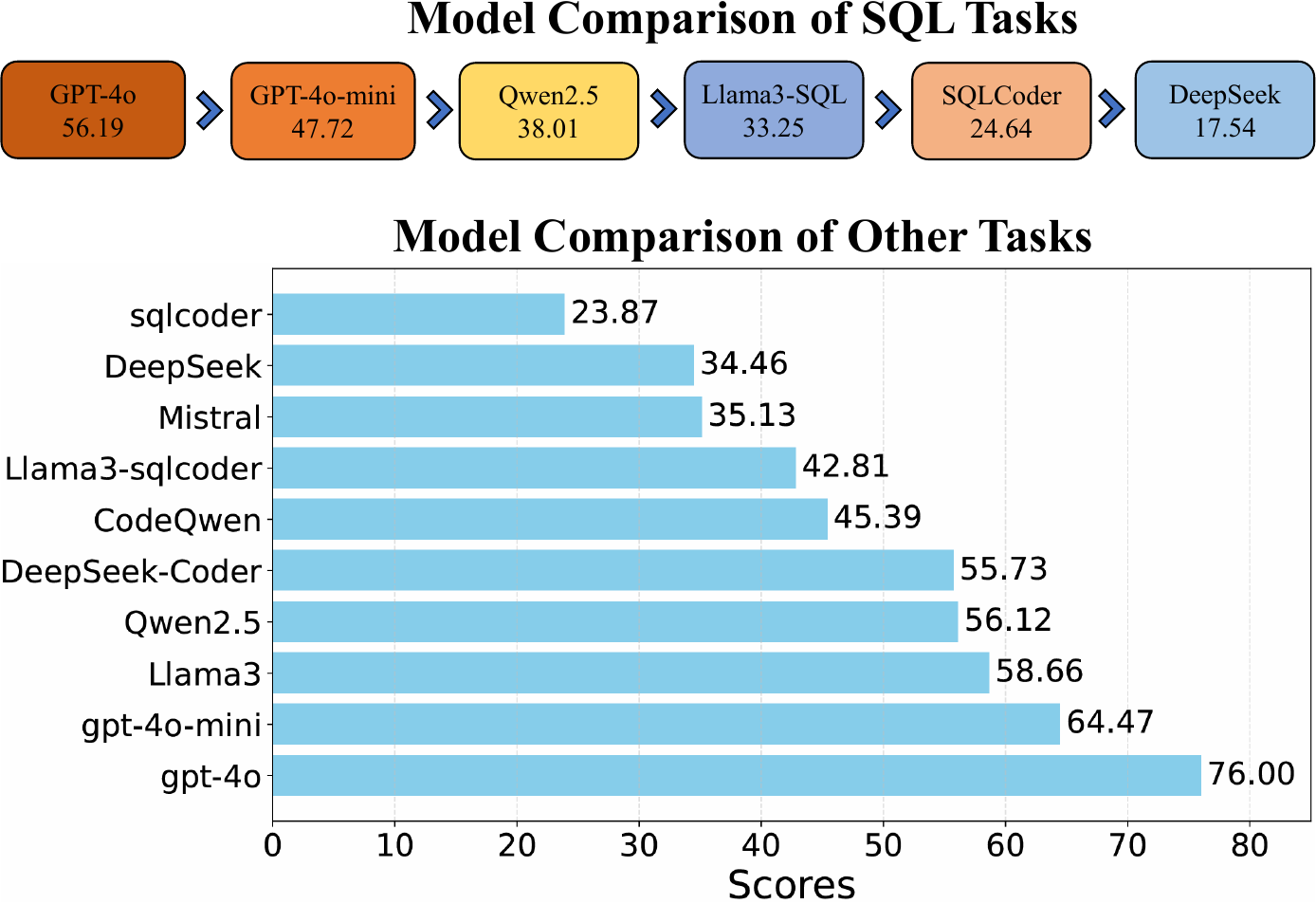}
    \caption{Ranking of models on SQL generation tasks and other (non-SQL generation) tasks.}
    \label{fig:rank}
    \vspace{-0.5em}
\end{figure}

Existing benchmarks are based on SQL generation tasks, such as Text-to-SQL, and do not address other types of database tasks. For further analysis, in Figure \ref{fig:rank}, we test the model's performance on non-SQL generation tasks (e.g., design, maintenance). Among them, SQLCoder performs poorly on other tasks compared to its performance on SQL generation tasks, which is due to the fine-tuning on SQL data affecting its capabilities on other tasks.

\section{Related Work}
\paragraph{LLM Applications in Database Tasks} Large language models (LLMs) have enabled significant advances in a variety of fields, including database management \cite{llm_db_manage,llm_db_2}. 
For example, LLMs have been employed for Text-to-SQL tasks, where they generate SQL queries based on natural language descriptions. A range of LLM-based Text-to-SQL methods have been proposed \cite{text2sql_survey,SQLizer,xu2017sqlnet}, showing substantial progress on this task. Additionally, LLMs have been used in database maintenance \cite{db_gpt}, a domain where recent work has contributed to advancements. As LLMs become more widely used in the database field, the challenge of evaluating their database capabilities has grown, necessitating a comprehensive benchmark.

\paragraph{Existing Database Benchmarks} Several benchmarks have been proposed to evaluate LLMs in the context of database-related tasks. SQLStorm \cite{schmidt2025sqlstorm}, BIRD \cite{BIRD} and Spider \cite{spider,spider2} are Text-to-SQL benchmarks designed to assess a model’s ability to generate SQL queries from natural language questions. They are widely used to evaluate the performance of general models in query generation, primarily focusing on execution accuracy (Execution Accuracy, EX). Additionally, Database QA benchmarks \cite{db_qa} have been introduced to assess the richness of models' database domain knowledge. However, both benchmarks are limited to a single task and fail to comprehensively assess other database-related tasks such as schema design, database debugging, or maintenance. We fill this gap by proposing a more comprehensive evaluation benchmark that covers the entire database lifecycle \cite{db_lifecycle}, including design, implementation, debugging, and maintenance.

\section{Conclusion}
We presented \textbf{DBLifeBench}, a pioneering benchmark that shifts the evaluation paradigm from isolated SQL generation to holistic database lifecycle management. By assessing capabilities across Design, Implementation, Operation, Debugging, and Maintenance, we exposed the limitations of current specialized models, which often sacrifice general reasoning for syntax proficiency. Additionally, our \textbf{Progressive-Text2SQL} task demonstrates that structured, step-by-step reasoning is key to solving complex database problems. DBLifeBench provides the community with a rigorous standard to guide the development of the next generation of LLMs-true Autonomous Database Agents.

\section*{Limitations}
DBLifeBench, while comprehensive, has some limitations. It is primarily designed for LLMs, which may not fully capture the performance of non-LLM approaches. Additionally, some tasks, particularly in design and maintenance, may be challenging to define and evaluate accurately. Lastly, the current benchmark focuses exclusively on textual modalities and does not address multimodal scenarios, such as interpreting visual performance charts. Future work will aim to incorporate multimodal inputs to broaden the scope of evaluation.

\bibliography{custom}

\clearpage
\appendix
\section{Appendix}
\label{sec:appendix}

\subsection{Examples}
\label{sec:appendix_example}
\begin{lstlisting}[language=MyLang, caption={Example of Design.}]
# Input
Requirements Analysis:
1. Purpose
2. Key Requirements
3. Non-Functional Requirements
...
# Output
Table 1: frpm
CDSCode, TEXT, Primary Key
...
Table 2: satscores
cds, TEXT, Primary Key
...
\end{lstlisting}
\begin{figure*}[t]
    \centering
    \includegraphics[width=\linewidth]{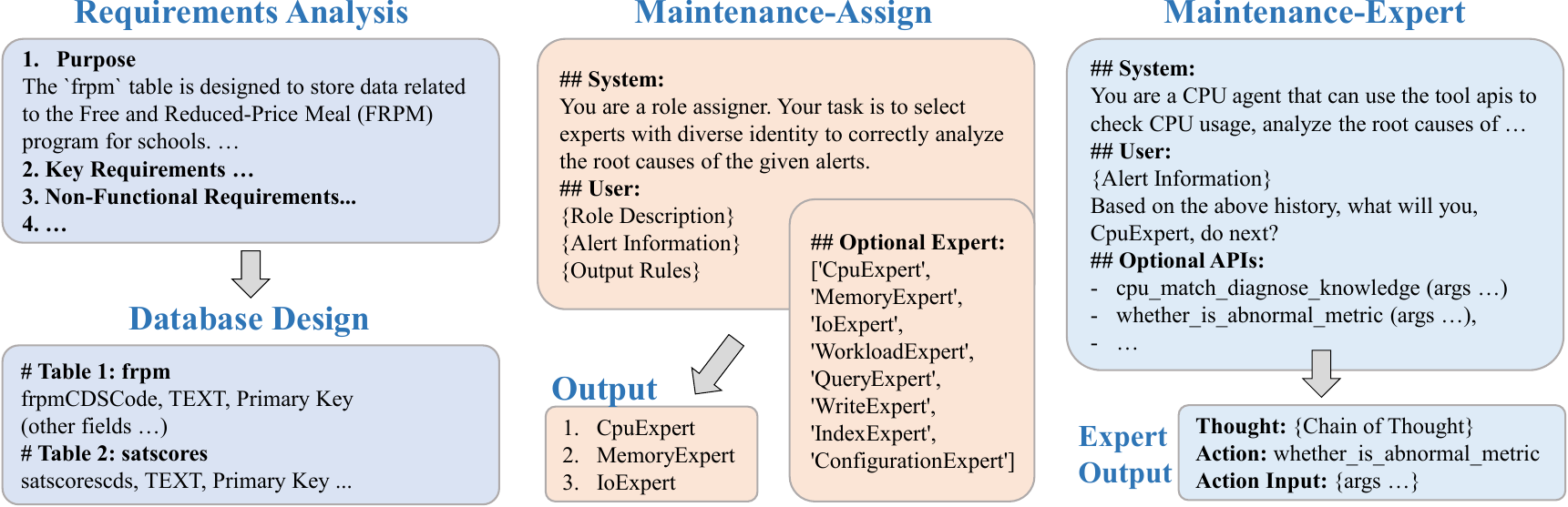}
    \caption{Illustration of Design and Maintenance}
    \label{fig:example}
\end{figure*}

\begin{table*}[t]
  \centering
    \begin{tabular}{lcccccc}
    \toprule
    \multirow{2}[0]{*}{Model} & \multicolumn{3}{c}{Text2SQL} & \multicolumn{3}{c}{P-Text2SQL} \\
    \cmidrule(lr){2-4}
    \cmidrule(lr){5-7}
          & \multicolumn{1}{c}{v1} & \multicolumn{1}{c}{v2} & \multicolumn{1}{c}{v3} & \multicolumn{1}{c}{v1} & \multicolumn{1}{c}{v2} & \multicolumn{1}{c}{v3} \\
    \midrule
    GPT-4o & 19.25  & 18.00  & 20.50  & 54.33  & 49.19  & 50.33  \\
    Llama3 & 13.75  & 11.75  & 14.75  & 42.67  & 40.45  & 48.33  \\
    Mistral & 9.50  & 9.25  & 8.75  & 30.00  & 22.98  & 30.00  \\
    DeepSeek & 11.75  & 10.00  & 12.00  & 36.33  & 30.74  & 33.33  \\
    Qwen2.5 & 14.25  & 14.00  & 17.00  & 49.33  & 41.42  & 49.67  \\
    ChatGLM-4 & 16.75  & 15.00  & 15.00  & 37.67  & 37.86  & 41.00  \\
    DeepSeek-Coder & 15.00  & 13.00  & 17.25  & 46.00  & 44.01  & 50.67  \\
    llama3-sqlcoder & 12.00  & 11.75  & 15.00  & 49.00  & 41.42  & 53.00  \\
    sqlcoder-7b-2 & 8.75  & 8.00  & 8.75  & 38.00  & 32.04  & 33.67  \\
    Qwen2.5-Coder-7B-Instruct & 9.00  & 8.25  & 11.25  & 36.33  & 33.01  & 38.00  \\
    
    \bottomrule
    \end{tabular}%
  \caption{Comparison of the model's performance across different schemas.}
  \label{tab:data_model}%
\end{table*}%

\begin{lstlisting}[language=MyLang, caption={Example of Implementation.}]
# Input
{results of design phase}
# Output
Table 1: CREATE TABLE frpm (
    CDSCode TEXT PRIMARY KEY,
    AcademicYear TEXT,
    ...
);
Table 2: CREATE TABLE satscores (
    cds TEXT PRIMARY KEY,
    ...
);
\end{lstlisting}

\begin{lstlisting}[language=MyLang, caption={Example of Text2SQL.}]
# Input
## Simple DDL:
albums (AlbumId, Title, ArtistId)
artists (ArtistId, Name)
...
## Question:
Retrieve the album titles and the corresponding artist names.
# Output
SELECT Title, Name FROM albums INNER JOIN artists USING (ArtistId);
\end{lstlisting}

\begin{lstlisting}[language=MyLang, caption={Example of P-Text2SQL.}, label=exp:P-Text2SQL]
# Input
Simple DDL: ...
Nodes:
{
    "id": 1,
    "sql statement": "SELECT ID FROM instructor WHERE name = 'David Brown'",
    "nl": "Retrieve the ID of the instructor whose name is 'David Brown'."
},
{
    "id": 2,
    "sql statement": "SELECT course_id, sec_id, semester, year FROM section WHERE building = 'Smith' AND room_number = '101'",
    "nl": "Retrieve the course ID, section ID, semester, and year for all sections held in room 101 of the 'Smith' building."
}
Edges:
{
    "source": 1,
    "target": 3
},
{
    "source": 2,
    "target": 3
}
"question":
Update the instructors of all courses held in room 101 of the 'Smith' building to 'David Brown'.
# Output
UPDATE teaches SET ID = (SELECT ID FROM instructor WHERE name = 'David Brown') WHERE (course_id, sec_id, semester, year) IN (SELECT course_id, sec_id, semester, year FROM section WHERE building = 'Smith' AND room_number = '101')
\end{lstlisting}

\begin{lstlisting}[language=MyLang, caption={Example of SQL-Debugging.}]
# Input
## Simple DDL: ...
## Wrong SQL:
SELECT T1.Phone FROM satscores AS T1 INNER JOIN schools AS T2 ON T1.cds = T2.CDSCode WHERE T1.AvgScrRead > 1500 ORDER BY T1.NumTstTakr DESC LIMIT 1
## Error Info:
OperationalError('no such column: T1.Phone')

# Output
SELECT T2.Phone FROM satscores AS T1 INNER JOIN schools AS T2 ON T1.cds = T2.CDSCode ORDER BY T1.NumGE1500 DESC LIMIT 1

\end{lstlisting}

\begin{lstlisting}[language=MyLang, caption={Example of Maintenance-Assign.}]
# Input
## System:
You are a role assigner. Your task is to select experts with diverse identity to correctly analyze the root causes of the given alerts.
## User:
- Role Description
You are the leader of a group of diagnosis experts. Currently, the database in your company may meet problems. The anomaly alert is:
During the inspection, it was found that from 2023-10-15 11:39:49 to 2023-10-15 11:40:49, the database's CPU usage was above the threshold of ...

Now you need to select experts with diverse identity to correctly analyze the root causes of the given alert. The names of available experts are:
['CpuExpert', 'MemoryExpert', 'IoExpert', 'WorkloadExpert', 'QueryExpert', 'WriteExpert', 'IndexExpert', 'ConfigurationExpert']

# Output
1. CpuExpert
2. MemoryExpert
3. IoExpert
\end{lstlisting}

\begin{lstlisting}[language=MyLang, caption={Example of Maintenance-Expert.}]
# Input
## System:
You are a CPU agent that can use the tool apis to check CPU usage (whether_is_abnormal_metric), analyze the root causes of high CPU usage using the metrics, queries and knowledge gained from (cpu_match_diagnose_knowledge), and give optimization solutions (e.g., optimize_index_selection, enable_or_disable_nestloop_operator).
## User:
You are in a company whose databases meet an anomaly. The anomaly's start_time is ${start_time} and end_time is ${end_time}. The anomaly alert is: {Alert ...}

- During diagnosis, you have access to the following tools:
{Optional APIs...}

{Database history}

Based on the above history, what will you, CpuExpert, do next?
# Output
Thought: Now that I have obtained the start and end time of the anomaly, I will check whether the CPU usage is abnormal during that time period using the `whether_is_abnormal_metric` tool.

Action: whether_is_abnormal_metric
Action Input: {"start_time": "2023-10-15 21:01:49", "end_time": "2023-10-15 21:02:49", "metric_name": "cpu_usage"}

Please execute the above action and provide the result.
\end{lstlisting}

\subsection{Graph Construction of Progressive-Text2SQL}
\label{apd:p-text2sql}
The following is the prompt for constructing graph data in P-Text2SQL, with the \textit{Examples} provided in Listing \ref{exp:P-Text2SQL}.

\begin{lstlisting}[language=python, caption={Prompt for Graph Construction.}]
    prompt = (
        "The goal is to construct a directed graph representation from a given sqlite SQL statement to represent the gradual advancement of functional implementation and eventually complete all requirements."
        "Each node in the graph represents a SQL statement, which is a subsequence of a given SQL statement, and the edge represents the execution order. "
        "Subsequent nodes are the progression of the previous node. ( please attention the last node should be the original given SQL statement. )"
        "This graph should faithfully reflect the topological execution order of SQL statements to achieve requirements. "
        "The following includes three cases, including Sql, and Graph.\n\n"

        "## Example 1\n"
        f"{example_template_1}\n\n"
        "## Example 2\n"
        f"{example_template_2}\n\n"
        "## Example 3\n"
        f"{example_template_3}\n\n"
        "Please build the graph according to the following Sql. Please note that just output the final graph. Do not include any other superfluous descriptions.\n\n"
        "# Sql:\n"
        f"{raw_sql}\n"
        "# Graph:\n")
\end{lstlisting}

\end{document}